\documentclass[11pt]{article}

\include{epsf}
\usepackage{latexsym}

\newcommand{\vx}{\ensuremath{\mathbf{x}} }
\newcommand{\vy}{\ensuremath{\mathbf{y}} }
\newcommand{\vmu}{\ensuremath{\mathbf{\hat{\mu}}} }
\newcommand{\pfmeas}{\ensuremath{{\cal D} \phi^* {\cal D} \phi} }
\newcommand{\pacc}{\ensuremath{{\cal P}_{\rm acc}} }
\newcommand{\capt}[1]{\begin{minipage}[h]{12cm} 
                      \caption[]{ \small #1} 
                      \end{minipage}}

\begin{document}

\title{
       Accelerating the Hybrid Monte Carlo algorithm \\ 
       with ILU preconditioning.}
\author{{\bf M.~J.~Peardon\thanks{
           Present address: School of Mathematics, 
                            Trinity College, Dublin 2, Ireland.
                            {\tt mjp@maths.tcd.ie} },
        }\\
        Dept.~of Physics, University of California, San Diego,\\
        La Jolla, California 92093-0319, USA.\\ \\
        \centerline{Preprint UCSD/PTH 00-17}
       }
\date{$24^{\rm th}$ November, 2000}
\maketitle 
\abstract{The pseudofermion action of the Hybrid Monte Carlo (HMC) algorithm 
for dynamical fermions is modified to directly incorporate Incomplete LU 
(ILU) factorisation. This reduces the stochastic noise and allows a larger
molecular dynamics step-size to be taken, cutting the computational cost.
Numerical tests using the two-flavour Schwinger model are presented, where 
a two-step ILU preconditioning of the even-odd fermion matrix allows the
step-size to be increased by a factor of two over the standard even-odd
formulation.}

\section{Introduction}

Monte Carlo integration of the partition function of QCD with light quarks 
remains a computationally demanding task.
At present, Hybrid Monte Carlo (HMC) \cite{Duane:1987de} is amongst the most 
widely used algorithms for generating an ensemble of gauge field 
configurations with the dynamical QCD probability distribution. This exact 
algorithm combines molecular dynamics 
evolution in a fictitious simulation time, with a Metropolis test to 
ensure detailed balance.
The effects of dynamical Wilson quark fields are 
introduced using gaussian-distributed ``pseudofermion'' fields and most of the
computational effort goes into inverting the
fermion matrix at each step of the molecular-dynamics trajectory. As a result, 
attention has focussed on improving the algorithms for this large, sparse 
matrix inversion. 

ILU preconditioning schemes are commonly used to accelerate iterative 
inverters. For an interaction matrix, this translates into first
ordering the sites on the lattice, then decomposing the matrix into 
upper and lower components. The upper segment couples a site only to those 
neighbours with a higher ordering index, and similarly the lower matrix
couples to the lower-ordered sites. The preconditioning matrices are then
constructed from these two terms. 
In one highly efficient inversion method for parallel machines, 
the SSOR scheme \cite{Fischer:1996th}, a ``locally-lexicographic'' ILU 
preconditioning is used and the fermion matrix is subsequently inverted using 
BiCGStab. In Ref.~\cite{Fischer:1996th}, the commonly used even-odd (or
``red-black'') scheme was recognised as an ILU decomposition, but not the 
optimal one.  Even-odd (eo) preconditioning of the pseudofermion coupling matrix
has also been used \cite{Gupta:1989kx} to reduce the stochastic noise in 
molecular dynamics evolution, and leads to an increase in the acceptance rate 
of the Metropolis test. In this paper, it is first noted that, following the 
even-odd example, any ILU preconditioning can be applied directly to the matrix 
appearing in the pseudofermion action. Beyond this, a simple two-step 
scheme is presented, where the matrix is first even-odd preconditioned and
the sites on one sub-lattice are subsequently ordered and ILU-factorised again.
The global-lexicographic ordered version of this two-step scheme was proposed
in Sec.~4 of Ref.~\cite{deForcrand:1997ck}.  In this work, other ordering
schemes are tested. The two-step method leads to further improvements in the 
solver convergence rate and when used in the pseudofermion action, the HMC 
algorithm performance is also significantly enhanced. 

The paper is organised as follows; Sec.~\ref{sec:theory} 
implements ILU preconditioning in the pseudofermion action and presents the 
two-level scheme, then in Sec.~\ref{sec:simulations}, these
algorithms are tested in simulations of the Schwinger model. Here, using the
two-step ILU preconditioned even-odd (eo-ILU) fermion matrix, a performance
improvement of a factor of about two is found compared to the standard even-odd
preconditioning. Sec.~\ref{sec:implementations} briefly discusses
application of the method in large-scale simulations on parallel computers and  
which use improved fermion actions.

\section{Preconditioning the pseudofermions in HMC. \label{sec:theory}}

  For simulations of a gauge theory with two degenerate flavours of dynamical 
fermions, the partition function is
\begin{equation}
  Z = \int\!{\cal D}U \;\;\; \det M ^2
       e^{-S_g}, 
\end{equation}
with $S_g$ the lattice Yang-Mills discretisation and $\det M$ the determinant 
of the fermion matrix. 
In the HMC algorithm, this determinant is re-expressed as a gaussian 
integral over new bosonic degrees of freedom; the ``pseudofermions''
\begin{equation}
  \det M^\dagger M \equiv \int\!\pfmeas \;\;\;
                      e^{-\phi^* \left[M^\dagger M\right]^{-1} \phi}.
\end{equation}
Notice that to ensure the gaussian integral is well defined, the number of 
fermion flavours simulated must be even ($N_f=2$ is assumed throughout this
work) and the $\gamma_5$-hermiticity
property of the Wilson fermion matrix has been used; ie
\begin{equation}
  M^\dagger = \gamma_5 M \gamma_5, 
      \mbox{\hspace{5ex}hence\hspace{5ex}} \det M = \det M^\dagger.
  \label{eqn:gamma5herm}
\end{equation}
Then the partition function to be
simulated is written as 
\begin{equation}
  Z = \int\!{\cal D}U \pfmeas \;\;\;
       e^{-S_g -\phi^* \left[M^\dagger M\right]^{-1} \phi}. \label{eqn:Z}
\end{equation}
The Hybrid Monte Carlo algorithm generates a new element in the
sequence of configurations 
in two stages. First, a fictitious continuous
time variable, $\tau$ is introduced along with canonical momentum variables,
$p_\mu(\vx)$ conjugate to each of the gauge degrees of freedom, $U_\mu(\vx)$. A 
hamiltonian describing dynamics in the new time coordinate is introduced,
\begin{equation}
  H = \frac{1}{2} p^2 + S_g +\phi^* \left[M^\dagger M\right]^{-1} \phi,
  \label{eqn:H}
\end{equation}
and the system is evolved in $\tau$ using a reversible, area-preserving
integration scheme such as leap-frog. Since the integration scheme is inexact,
the hamiltonian is not conserved and a subsequent stochastic step is added to
compensate. After some time interval, the new
configuration is proposed as an element of the ensemble and accepted or
rejected according to the Metropolis test on the change in the hamiltonian; 
\begin{equation}
  P_{\rm acc} = \min(1,e^{-\delta H}). 
\end{equation}
The high computational cost of these dynamical fermion simulations arises as
calculating the force term acting on the conjugate momenta at each leap-frog 
step requires two inversions of the fermion matrix, $M$. 

  An Incomplete LU (ILU) factorisation of $M$ was demonstrated as
an efficient means of accelerating inversion \cite{Oyanagi,Fischer:1996th} and 
has been used successfully in large-scale production runs by {\it eg.} the 
SESAM collaboration \cite{Lippert:1998qy}. The ILU factorisation preconditions 
the fermion matrix, $M$ by left and right multiplication with two readily 
invertible matrices; 
\begin{equation}
  \bar{M} = (I-L)^{-1} M (I-U)^{-1}
    \label{eqn:Mbar-def}
\end{equation}
where $L$ and $U$ are the lower and upper parts of the Wilson hopping term.
Defining $L$ and $U$ first requires the sites on the lattice, \vx are assigned 
an integer index, $s(\vx)$ then the site ordering is defined as
\begin{equation}
  \vy > \vx \mbox{\hspace{3ex}if\hspace{3ex}} s(\vy) > s(\vx),
\end{equation}
with $\vy < \vx$ defined similarly. The lower part of the Wilson matrix is then 
\begin{equation}
  L_{\vx\vy} = \left(
      \begin{array}{cr}
      \kappa \sum_{\mu} 
      U_\mu(\vx) (1-\gamma_\mu) \delta_{\vy,\vx+\vmu} + 
      U^{\dagger}_\mu(\vx-\vmu) (1+\gamma_\mu) \delta_{\vy,\vx-\vmu}
      & \mbox{when }\vy<\vx \\
      0 
      & \mbox{otherwise}
      \end{array}
      \right.
\end{equation}
and $U$ is defined similarly for sites where $\vy>\vx$. 
The full Wilson matrix is then equivalent to 
\begin{equation}
  M = I-L-U. \label{eqn:M-def}
\end{equation}
Matrix inversion is
accelerated since the new matrix, $\bar{M}$ is better conditioned than $M$. The
preconditioning matrices, $(I-L)$ and $(I-U)$ are easily inverted by either 
forward or backward substitution respectively. The $\gamma_5$-hermiticity 
({\it cf.} Eqn.~\ref{eqn:gamma5herm}) of the preconditioned matrix is 
preserved as 
\begin{equation}
  \bar{M}^\dagger = \gamma_5 \bar{M} \gamma_5 
     \mbox{\hspace{5ex}since\hspace{5ex}} 
     (I-L)^\dagger = \gamma_5 (I-U) \gamma_5.
\end{equation}
Matrix-vector operations proceed efficiently via the ``Eisenstat trick''
\cite{Eisenstat,Fischer:1996th}; using Eqns.~(\ref{eqn:Mbar-def}) and 
(\ref{eqn:M-def}), $\bar{M}$ is re-written as 
\begin{equation}
\bar{M} = (I-U)^{-1} + (I-L)^{-1} \left(I - (I-U)^{-1}\right),
\end{equation}
and the matrix operation is reduced to a backward substitution followed by a 
forward one. This requires approximately the same number of floating
point operations as the original Wilson matrix-vector product.

While the introduction of the pseudofermions to model the fermion
determinant makes the molecular dynamics of the hamiltonian tractable, it also
introduces extra randomness into this evolution. Simulations demonstrate that 
the Metropolis acceptance rate is higher with the better-conditioned even-odd 
matrix. 
For this study, first note that any ILU decomposition can be applied 
directly to the determinant of dynamical fermion simulations;
\begin{equation}
  \det \bar{M} = \det (I-L)^{-1} \det (I-L-U) \det (I-U)^{-1} = \det M,
\end{equation}
since $\det (I-L) = \det (I-U) = 1$. From this identity, the 
fermion determinant can be simulated using pseudofermions coupled via the 
preconditioned matrix, $\bar{M}$
\begin{equation}
  Z = \int\!{\cal D}U \pfmeas \;\;\;
       e^{-\phi^* \left[\bar{M}^\dagger \bar{M}\right]^{-1} \phi}.
\end{equation}
The HMC algorithm can be applied to this new pseudofermionic partition
function and the
better conditioning of the matrix should lead to an improvement
in the acceptance rate. The extent of this improvement will depend on the site
ordering used, while physical expectation values computed on the ensemble will
not. This will be tested in Sec. \ref{sec:simulations}

\subsection{Molecular dynamics for the preconditioned action.
  \label{sec:molecular}}

The new hamiltonian generating the molecular dynamics is
\begin{equation}
  \bar{H} = \frac{1}{2} p^2 + S_g 
            +\phi^* \left[\bar{M}^\dagger \bar{M}\right]^{-1} \phi,
  \label{eqn:HMC-hamiltonian}
\end{equation}
In order to perform the molecular-dynamics updates, the force term acting on 
the momenta of the hamiltonian of Eqn.~(\ref{eqn:HMC-hamiltonian}) must be 
determined. To demonstrate that the force term for the ILU preconditioned 
matrix, $\bar{M}$ is readily implemented, first note that the force arising 
from the gauge action is trivially left unchanged and consider the term arising 
from the pseudofermionic action alone,
\begin{equation}
  \bar{S}_{\rm pf} = \phi^* \left[\bar{M}^\dagger \bar{M}\right]^{-1} \phi.
\end{equation}
The derivative of this action with respect to $\tau$, keeping $\phi$ fixed is 
\begin{equation}
  \frac{d\bar{S}_{\rm pf}}{d\tau} = 
      -\phi^* \left[\bar{M}^\dagger \bar{M}\right]^{-1} 
         \left[\frac{d\bar{M}^\dagger}{d\tau} \bar{M}
              + \bar{M}^\dagger\frac{d\bar{M}}{d\tau}\right]
      \left[\bar{M}^\dagger \bar{M}\right]^{-1} \phi.
\end{equation}
Introducing the auxiliary fields, $Y = \bar{M}^{\dagger-1} \phi$ and 
$X = \bar{M}^{-1}Y$, this becomes
\begin{equation}
  \frac{d\bar{S}_{\rm pf}}{d\tau} = -Y^* \frac{d\bar{M}}{d\tau} X -
                                    X^* \frac{d\bar{M}^\dagger}{d\tau} Y.
\end{equation}
At each leap-frog step, the fields $Y$ and $X$ must be recomputed; this is the
section of the update requiring most of the computational effort. 
Note that the inversion proceeds more rapidly than in the original HMC
algorithm as $\bar{M}$ is better conditioned. The derivative of $\bar{M}$ can
be expanded after two new auxiliary fields are introduced; they are
\begin{equation}
  Y_L = (I-L^\dagger)^{-1} Y \mbox{\hspace{4ex}and\hspace{4ex}} 
                 X_U = (I-U)^{-1} X.
\end{equation}
Note that constructing these fields requires only two backward 
substitutions ($L^\dagger$ is an upper-diagonal matrix), rather than an 
iterative method and so is an insignificant overhead compared to re-evaluating 
$X$ and $Y$.
With these new fields, and using the definition of $\bar{M}$ in 
Eqn.~(\ref{eqn:Mbar-def}), the derivative becomes
\begin{equation}
  \frac{d\bar{S}_{\rm pf}}{d\tau} = 
     -\left( 
       Y_L^* \frac{dM}{d\tau} X_U 
     + \phi^*\frac{dU}{d\tau} X_U
     + Y_L^* \frac{dL}{d\tau} Y +\mbox{ h.c.} \right). \label{eqn:md-end}
\end{equation}
Now derivatives of the original Wilson matrix, $M$ and its upper and lower 
sections appear in Eqn.~(\ref{eqn:md-end}) and so calculating the
relevant force terms proceeds straightforwardly from here. 

\subsection{Even-odd (eo) and two-stage (eo-ILU) preconditioning 
   \label{sec:eoilu}}

In Ref. \cite{Fischer:1996th}, it was demonstrated that the commonly used 
even-odd preconditioning of the fermion matrix can be written as an ILU
decomposition where the ordering function, $s(\vx)$ is simply 0 on even
lattice sites and 1 on odd sites. There is no ordering ambiguity for the
sites on the even or odd sub-lattices since $M$ does not contain hopping terms
that directly couple sites on the same sub-lattice. For this example, the
inversion of $(I-L)$ and $(I-U)$ can be written explicitly. 

The preconditioned matrix is 
\begin{eqnarray}
  \bar{M} & = &
 \left(\begin{array}{cc} I_{oo}&0 \\  \kappa \Delta_{eo} & I_{ee}
      \end{array}\right)
 \left(\begin{array}{cc} I_{oo}&-\kappa\Delta_{oe}\\-\kappa\Delta_{eo}&I_{ee}
      \end{array}\right)
 \left(\begin{array}{cc} I_{oo}&\kappa\Delta_{oe}\\ 0 &I_{ee}
      \end{array}\right) \nonumber \\
        & = & 
 \left(\begin{array}{cc} I_{oo}&0 \\0& I_{ee}-\kappa^2 \Delta_{eo}\Delta_{oe}
      \end{array}\right). \label{eqn:Meo-def}
\end{eqnarray}
Pseudofermion degrees of freedom on the odd sites are completely decoupled 
from the gauge fields and can be discarded. The pseudofermions on
even sites are then coupled via the sub-matrix, $M_{ee} = I_{ee}-\kappa^2 
\Delta_{eo}\Delta_{oe}.$

Since $M_{ee}$ is ``ultra-local'' (the only non-zero elements of the matrix
couple sites within a small neighbourhood) and its elements can be written 
explicitly, it can be ILU preconditioned once again. The global lexicographic
version of this second level of preconditioning was investigated in Section 4 
of Ref.~\cite{deForcrand:1997ck}.  If a new ordering 
function, $s_e(\vx_e)$ is defined for sites on the even sub-lattice, then 
the (two-step) eo-ILU preconditioned matrix is 
\begin{equation}
  \bar{M}_{ee} = (I-L_{ee})^{-1} M_{ee} (I-U_{ee})^{-1},
\end{equation}
with
\begin{equation}
  L_{ee} = \kappa^2 \sum_{\vy_e<\vx_e} 
            \left. \Delta_{eo}\Delta_{oe}\right|_{\vx_e,\vy_e}. 
\end{equation}
As before, the matrix determinant is left unchanged by this second level of 
preconditioning; 
\begin{equation}
  \det \bar{M}_{ee} = \det M_{ee} = \det M.
\end{equation}
From here, a two-step preconditioned pseudofermion action can be 
implemented in the HMC algorithm, following a similar construction to
Sec.~\ref{sec:molecular}. The new pseudofermion action is 
\begin{equation}
  \bar{S}_{\rm e} = 
      \phi_e^* \left[\bar{M}_{ee}^\dagger \bar{M}_{ee}\right]^{-1} \phi_e,
\end{equation}
and, after introducing the auxiliary fields (on even sites only) 
$Y_e = \bar{M}_{ee}^{\dagger -1} \phi_e$ and $X_e = \bar{M}_{ee}^{-1} Y_e$ the
derivative with respect to $\tau$ becomes 
\begin{equation}
  \frac{d\bar{S}_{\rm e}}{d\tau}=
              -Y_e^* \frac{d\bar{M}_{ee}}{d\tau} X_e -
               X_e^* \frac{d\bar{M}_{ee}^\dagger}{d\tau} Y_e.
\end{equation}
Finally, after introducing the new fields $Y_{Le} = (I-L_{ee}^\dagger)^{-1} Y_e$
and $X_{Ue} = (I-U_{ee})^{-1} X_e$, the derivative is 
\begin{equation}
  \frac{d\bar{S}_{\rm pf}}{d\tau} = 
     -\left( 
       Y_{Le}^* \frac{dM_{ee}}{d\tau} X_{Ue}
     + \phi_e^* \frac{dU_{ee}}{d\tau} X_{Ue}
     + Y_{Le}^* \frac{dL_{ee}}{d\tau} Y_e +\mbox{ h.c.} \right).
          \label{eqn:md-eo-end}
\end{equation}
This expression contains derivatives of the original even-odd matrix and its
lower and upper components. The first term is efficiently computed using 
the reconstruction trick; $Y_L$ and $X_U$ are defined on odd sites as 
 $X_{Uo} = \kappa \Delta_{oe} X_{Ue}$ and 
 $Y_{Uo} = \kappa \Delta^\dagger_{oe} Y_{Ue}$
then 
\begin{equation}
   Y_{Le}^* \frac{dM_{ee}}{d\tau} X_{Ue} \equiv Y_L^* \frac{dM}{d\tau} X_U
\end{equation}
and so the force term is readily computed. Unfortunately no such
reconstruction can be used for the terms involving $U_{ee}$ and $L_{ee}$.
The reconstruction trick relies on the observation that the preconditioning
matrices for the even-odd decomposition can be computed explicitly (as in
Eqn.~\ref{eqn:Meo-def}). 
While these algebraically cumbersome force terms must be evaluated explicitly,
their evaluation is still a simple, local computational task.

\section{Testing the method: the Schwinger model\label{sec:simulations}}

 To investigate the benefits of preconditioning, simulations of the two-flavour 
Schwinger model were performed. This model is a $U(1)$ gauge theory in 
$1+1$ dimensions and is a convenient testing ground for algorithms intended 
for full QCD simulations as it is asymptotically free, confining and has a 
spontaneously broken chiral symmetry. 
The link variables are phases; $U_\mu(\vx) = \exp\; i \theta_\mu(\vx)$ and 
for this study, the compact $U(1)$ gauge action is used
\begin{equation}
   S_g  = \beta \sum_\Box 1 - \cos \theta_\Box,
   \label{eqn:s-gauge}
\end{equation}
with $\theta_\Box$ the sum of angles around the plaquette.

\subsection{ILU orderings \label{sec:ilu-order}}

\begin{figure}[h]
  \begin{center}
  
    \begin{minipage}[h]{12cm}
    \setlength{\unitlength}{0.38mm}
    \begin{picture}(330,120)(0,0)
      \setlength{\epsfxsize}{3.5cm}
      \put(0,20){\epsfbox{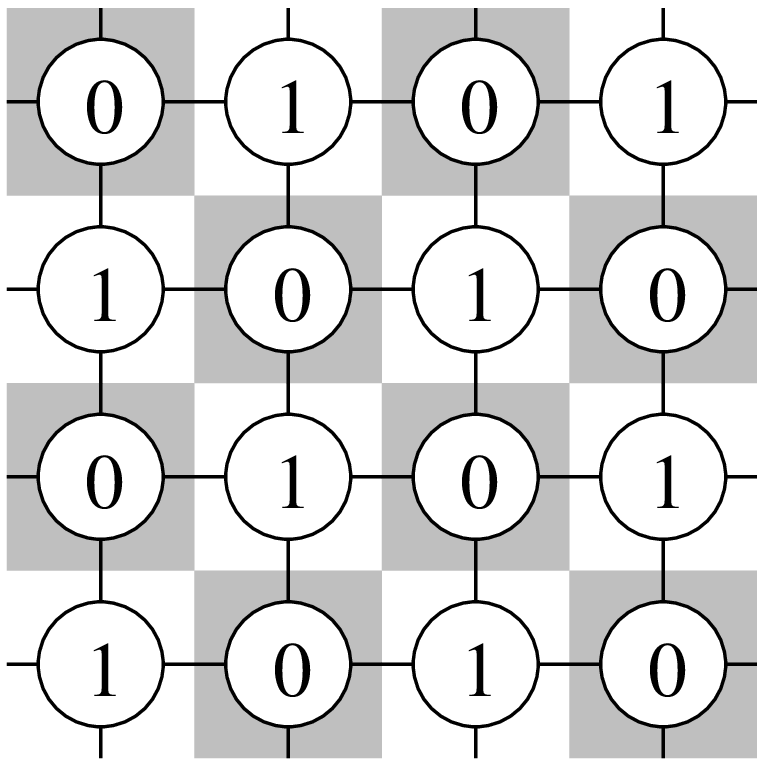}}
      \put(40,0){(a)}

      \setlength{\epsfxsize}{3.5cm}
      \put(110,20){\epsfbox{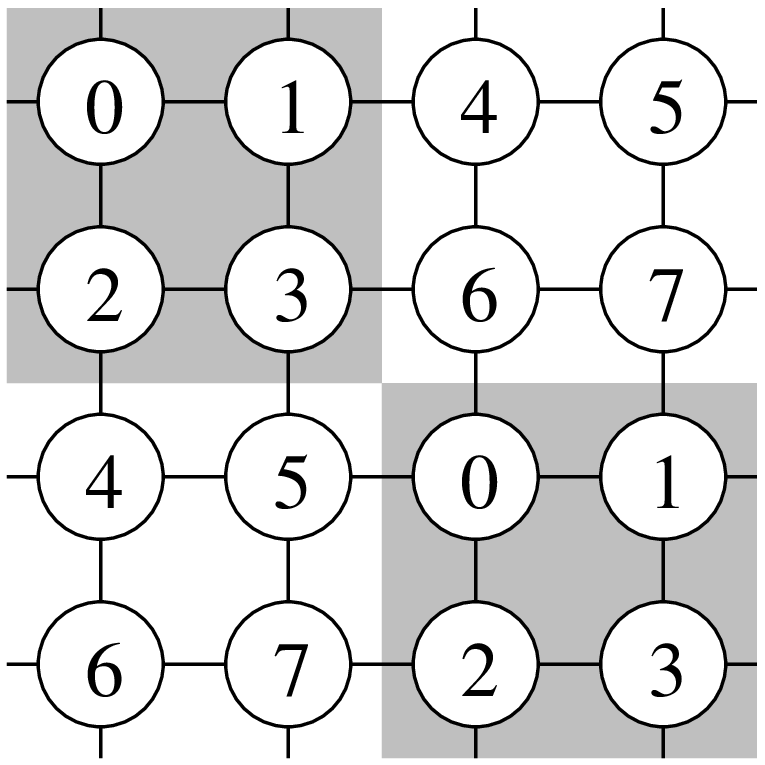}}
      \put(150,0){(b)}

      \setlength{\epsfxsize}{3.5cm}
      \put(220,20){\epsfbox{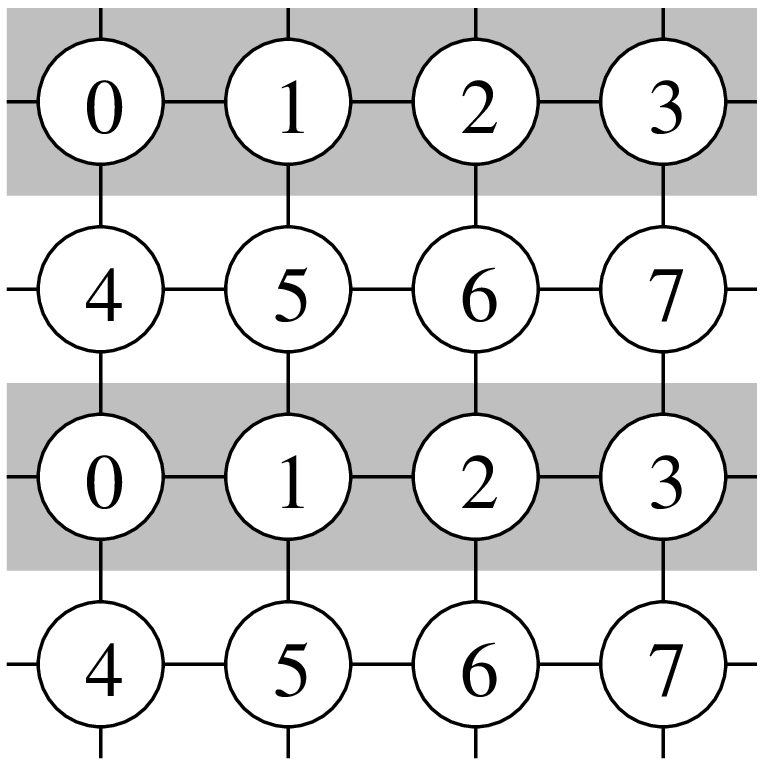}}
      \put(260,0){(c)}

    \end{picture}
    \caption[]{\small The index function, $s(x)$ for various lattice 
preconditionings. Grey shading denote areas that are lexicographically ordered. 
(a) is the standard even-odd preconditioning, denoted $ll_1$ (b) is the {\it 
locally lexicographic} scheme with block size 2, $ll_2$ and (c) is the {\it 
strip-lexicographic} scheme, $sl_1$. \label{fig:ilu-lattices} }

    \end{minipage}

  \end{center}
\end{figure}

Ref.~\cite{Fischer:1996th} noted that the performance of ILU 
preconditioned inversion depends on the choice of ordering of the 
lattice sites. To study this, the spectral properties of the matrix were
determined for a number of different schemes. The smallest and largest
eigenvalues (and hence the condition number) of the hermitian matrix, 
$\bar{Q}$ ($\equiv \gamma_5 \bar{M}$) were computed on an ensemble of 100 
dynamical $\beta=4.0, \kappa=0.26, 32\times 32$ lattices. The orderings tested 
were the standard even-odd checkerboard, a {\it locally lexicographic (ll)} 
scheme and a {\it strip-lexicographic (sl)} scheme. For the $(ll_n)$ ordering, 
the lattice is first decomposed into an even-odd checkerboard of $n \times n$ 
blocks then the sites in each block are indexed starting at the corner with 
smallest coordinates and progressing first along the $x$-axis until the end of 
the block is reached before moving onto the next $y$ value. Note that $(ll_1)$ 
is just the traditional even-odd indexing and $(ll_N)$, where $N$ is the 
lattice extent, denotes a global lexicographic ordering. The $(sl)$ ordering 
first breaks the lattice into $1 \times N$ strips and then the sites on each 
strip are indexed in order. Fig.~\ref{fig:ilu-lattices} illustrates these 
ordering schemes on a $4 \times 4$ lattice. 

\begin{table}[ht]
\begin{center}
\begin{tabular}{|c|c|c|c|c|}
\hline
Ordering               & $|\lambda|_{\rm min}$ & $|\lambda|_{\rm max}$ 
                       & ${\cal C}$       & $n_{\rm iter}$ \\
\hline
\hline
 None                  &    0.01768(48)      &   2.02173(11)
                       & 123.7(3.7)       &  501(11) \\
\hline
 eo ($ll_1$)           &    0.03511(94)      &   1.56235(76)
                       &  {\bf 48.1(1.4)} &  202(5) \\
  $ll_2$               &    0.0419(11)       &   1.9873(14)
                       &  51.2(1.5)       &  104(3) \\
  $ll_4$               &    0.0439(11)       &   2.8837(29)
                       &  70.7(2.2)       &  66.3(3) \\ 
  $ll_N$               &    0.0444(11)       &   4.706(15)
                       & 114.0(3.5)       &  {\bf 49.1(3)} \\ 
  $sl_1$               &    0.0443(12)       &   2.7254(23)
                       &  66.4(2.0)       &  104(1) \\ 
\hline
\end{tabular}
\capt{Spectral properties of $\bar{M}$ (and $\bar{Q}$) at $\beta=4.0, 
\kappa=0.26$ on a $32\times 32$ lattice for various ILU schemes. The labelling
of the ILU schemes is described in the text. Results for the schemes with 
optimal condition number and solver performance are in bold.
  \label{tab:spectral}}
\end{center}
\end{table}

Table~\ref{tab:spectral} presents the spectral properties of the hermitian 
fermion matrix, $\bar{Q}$ for a number of these schemes computed on 100
dynamical configurations. The eigenvalues of
$\bar{Q}$ with the largest and smallest magnitudes are given, along with their
ratio, ${\cal C}$ and $n_{\rm iter}$, the number of BiCGStab iterations 
required to invert the fermion matrix. All the eigenvalues of $\bar{Q}$ were 
computed using the Lanczos algorithm with reorthogonalisation. Their 
reliability was checked by testing $\det \bar{Q} = \det Q $ 
configuration-by-configuration. This relation was seen to hold to machine 
precision. 
Note that to ensure consistently accurate solutions in column 5, 
$|Mx-y|/|y| < 10^{-12}$ was used as the stopping criterion for the BiCGStab 
solver ({\it ie.} the reconstructed residual of $M$ rather than $\bar{M}$). 

These results confirm that the 
condition number of the preconditioned matrix and the solver performance is 
strongly dependent on the site ordering scheme. An unusual pattern 
emerges however; the preconditioning that leads to optimal solver performance 
(measured by the number of iterations required for convergence) is the global 
lexicographic scheme, $ll_N$ but the scheme with the lowest condition number is
the standard even-odd decomposition. These two optimal orderings are
highlighted in Table~\ref{tab:spectral}. The $ll_N$ inverter
out-performs the even-odd method by a factor of four in iterations.
\begin{figure}
  \begin{center}
    \begin{minipage}[h]{12cm} 
      \setlength{\epsfxsize}{11cm}
      \centerline{\epsfbox{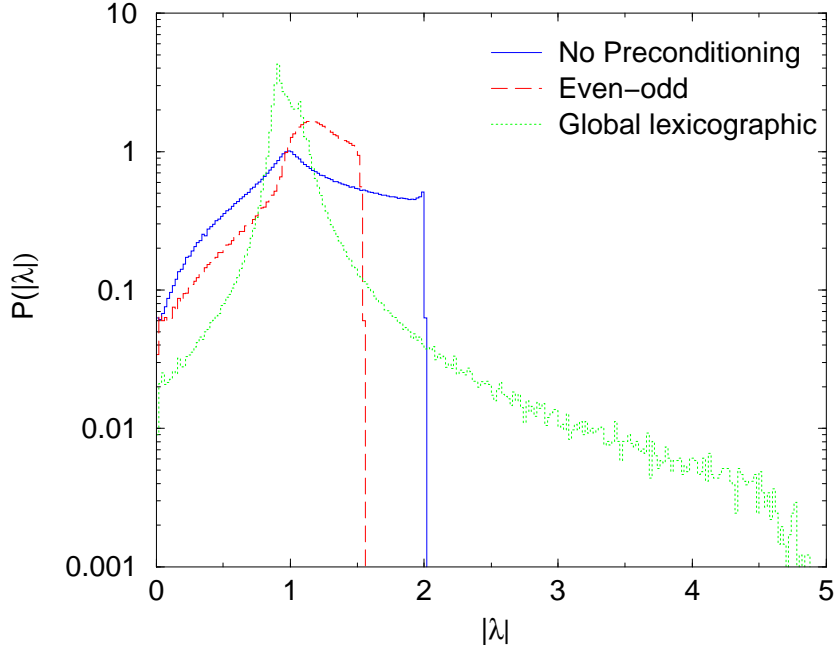}}
      \caption{The eigenvalue probability distribution of $Q$ and $\bar{Q}$.
         \label{fig:lambda}}
    \end{minipage}
  \end{center}
\end{figure}

The mis-match between the optimal ordering for better conditioning and for
inversion may be explained by closer examination of the eigenvalue spectrum.
Fig.~\ref{fig:lambda} shows the density of (the absolute value of) the 
eigenvalues of $Q$ without preconditioning and using even-odd and global
lexicographic ILU schemes. For the $(ll_N)$ scheme, most of the eigenvalues are
distributed close to unity while a small number lie in a long tail stretching
out to $|\lambda|=5$. This tail, while responsible for the
high condition number is sparsely populated and so easily handled by the 
BiCGStab algorithm. The even-odd matrix has a hard upper bound on its
eigenvalues and so is better conditioned, but these eigenvalues are broadly
distributed inside that band. Note also that the performance results in
Table.~\ref{tab:spectral} are for inversion of $M$, not $Q$. The spectrum of 
$M$ has eigenvalues that are generally complex and has not been computed here; 
solver performance may depend on the eigenvalue distribution in the complex 
plane. A heuristic argument for the better performance of the 
global-lexicographic ordering can be made by noting that each iteration of the 
solver couples every site on the lattice with every other site.

\subsection{ILU HMC algorithm performance \label{sec:ILU-HMC-perf}}

 The ordering schemes of Sec.~\ref{sec:ilu-order} were used to precondition the
matrix coupling pseudofermions in a set of dynamical fermion simulations on
lattices of size $32 \times 32$. The
gauge coupling was fixed to $\beta=4.0$ throughout. For the computations
of the $X$ and $Y$ fields, the BiCGStab algorithm was again used (the
stopping criterion was based on the residual of the preconditioned solution
vector, to reduce computational overheads). In
the molecular dynamics leap-frog step, the Sexton-Weingarten integrator 
scheme \cite{Sexton:1992nu} was implemented. Here, the force on the conjugate
momenta is separated into two parts; one arising from the pseudofermion
action and the other from the Yang-Mills term of 
Eqn.~(\ref{eqn:s-gauge}).
By interleaving the different momenta and gauge field updates, 
the step-size used for integrating the Yang-Mills force term, $d\tau_g$ is 
then made smaller than the step-size for the pseudofermionic force, $d\tau_f$. 
This isolates any changes in the update hamiltonian arising from the 
preconditioned fermion determinant. 
For all simulations, $d\tau_g = \frac{1}{4} d\tau_f$ was used.
This value was chosen in some short tests and subsequently 
found to be sufficient to remove the finite-step-size effects
of the gauge action in all runs. The molecular-dynamics trajectory length was 
selected randomly in the interval $\frac{1}{2} < \tau < 1\frac{1}{2}$ to 
ensure ergodicity.

Table~\ref{tab:wloop} shows the expectation values of the $1\times 1$ and 
$4 \times 4$ Wilson loops from simulations using the preconditioning orderings
of Sec.~\ref{sec:ilu-order}. All simulations were performed on a $32\times 32$
lattice, at $\beta=4.0, \kappa=0.2600$. Wilson loop averages agree within 
statistical uncertainties, as expected. Also in Table~\ref{tab:wloop} are the
acceptance probabilities of the final Metropolis test at the end of each
molecular dynamics trajectory and (for reference) the condition numbers, 
${\cal C}$ from Table.~\ref{tab:spectral} are included. While all the ILU 
preconditioned simulations outperform the standard HMC algorithm, there is 
little variation among the different orderings. The optimal schemes use 
even-odd and strip-lexicographic ordering. The global-lexicographic scheme, 
while the best ordering for inversion, is the worst for HMC performance. It is
possible that the higher condition number of $\bar{Q}$ in this scheme, due to 
the long tail of eigenvalues seen in the spectrum of Fig.~\ref{fig:lambda},
leads to instabilities in the molecular-dynamics evolution at a smaller 
step-size. Notice there is a correlation between the highest acceptance rates
and the lowest condition numbers. 
\begin{table}[t]
\begin{center}
\begin{tabular}{|c|c|c|c|c|c|}
\hline
Ordering            & $N_{\rm sweep}$ & 
            $\langle W(1,1)\rangle$ & $\langle W(4,4)\rangle$ 
                    & \pacc     & ${\cal C}$   \\
\hline
\hline
 None               & 5000         & 
              0.87348(41)             & 0.1800(31)            
                    & 0.120(11) & 123.7(3.7) \\
\hline
eo $(ll_1)$   & 10000        &
              0.87407(14)             & 0.18542(82)           
                    & 0.7310(74)& 48.1(1.4) \\
$ll_2$        & 1000         &
              0.87420(43)             & 0.1815(25)            
                    & 0.721(17) & 51.2(1.5) \\ 
$ll_4$        & 1000         &
              0.87416(34)             & 0.1834(27)            
                    & 0.691(16) & 70.7(2.2) \\ 
$ll_N$        & 1000         &
              0.87413(48)             & 0.1842(30)            
                    & 0.647(21) & 114.0(3.5) \\ 
$sl_1$        & 1000         &
              0.87360(37)             & 0.1826(27)            
                    & 0.745(14) & 66.4(2.0) \\ 
\hline
\end{tabular}
\capt{Wilson loop expectation values and acceptance probabilities from 
simulations using different preconditioning site ordering schemes. All
simulations were performed on $32\times 32$ lattices at $\beta=4.0,
\kappa=0.26$ with an MD step-size of $d\tau=\frac{1}{24}$.
\label{tab:wloop}}
\end{center}
\end{table}

\subsection{eo-ILU preconditioning}
 
\begin{figure}[h]
  \begin{center}
  
    \begin{minipage}[h]{12cm} 
    \setlength{\unitlength}{0.38mm}
    \begin{picture}(330,120)(0,0)
      \setlength{\epsfxsize}{3.5cm}
      \put(0,20){\epsfbox{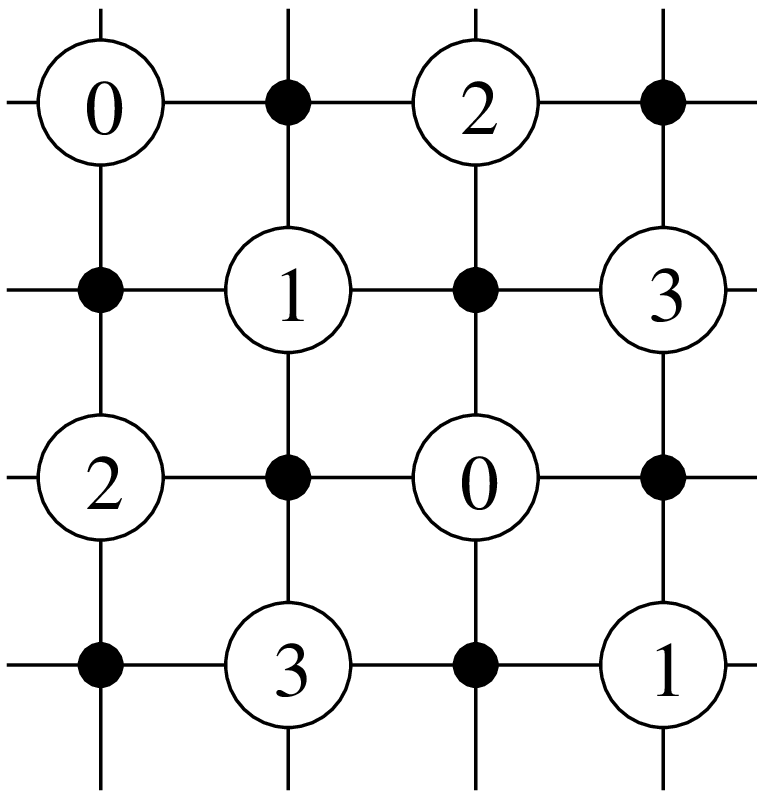}}
      \put(48,0){\makebox(0,0){(a)}}

      \setlength{\epsfxsize}{3.5cm}
      \put(110,20){\epsfbox{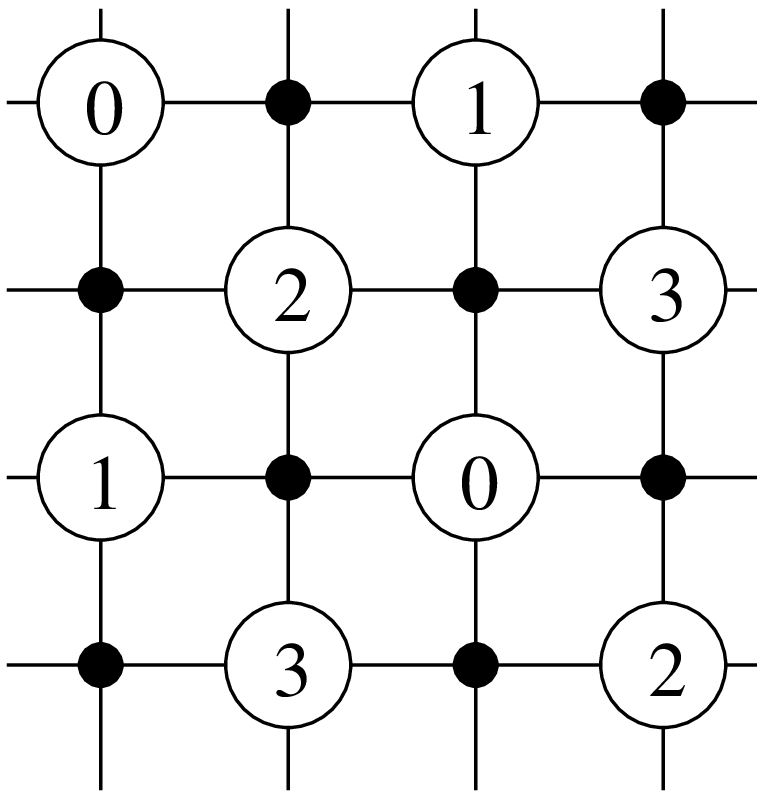}}
      \put(158,0){\makebox(0,0){(b)}}

      \setlength{\epsfxsize}{3.5cm}
      \put(220,20){\epsfbox{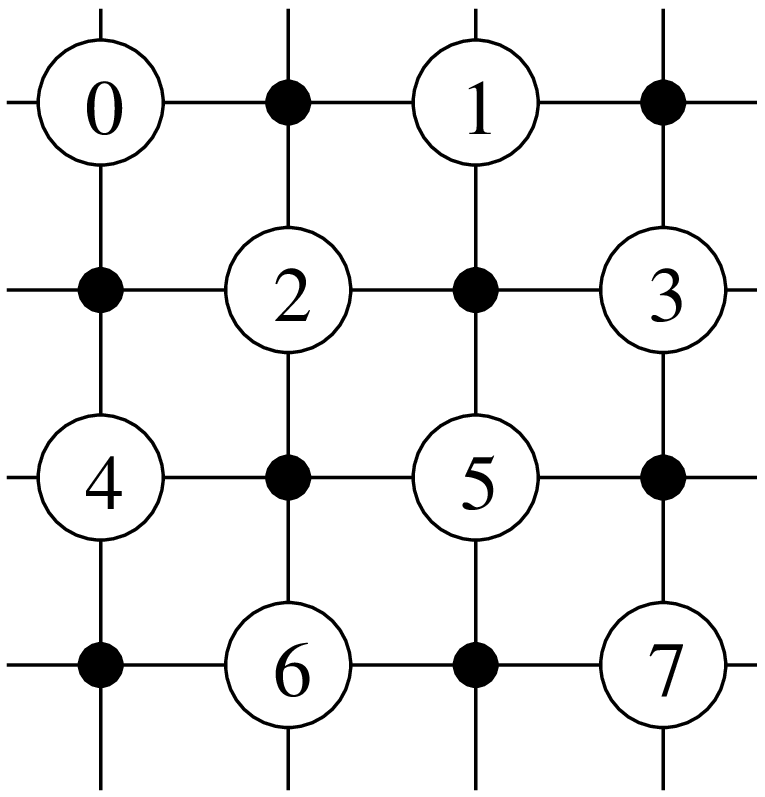}}
      \put(268,0){\makebox(0,0){(c)}}

    \end{picture}
    \caption[]{\small Two-step eo-ILU preconditioning schemes. (a) and (b)
are the two distinct local decompositions, (in two dimensions). In (c) the 
entire even sub-lattice is lexicographically ordered. 
\label{fig:eoilu-lattices} }

    \end{minipage}

  \end{center}
\end{figure}

Sec.~\ref{sec:eoilu} introduced a two-step preconditioning scheme (eo-ILU), 
where initially an even-odd decomposition of the fermion matrix is applied, 
followed by an ILU preconditioning. In the second step, an ordering scheme for 
the sites on the even (or odd) sites only is defined. In this section, 
the performance of the BiCGStab solver and HMC algorithm on eo-ILU matrices is
presented. Three ordering schemes are tested; the first is the standard
global-lexicographic scheme and the second two are the two local
preconditionings, labelled A and B, which are the extension of even-odd
preconditioning on the sub-lattice. In order to avoid ambiguity, four
``flavours'' of sub-lattices must be defined, since the even-odd matrix couples
the eight sites with off-set vectors $\{\pm1,\pm1\}, \{\pm2,0\}$ and 
$\{0,\pm2\}$. These three preconditionings are illustrated for a $4\times 4$
lattice in Fig.~\ref{fig:eoilu-lattices}. 

\begin{table}[ht]
\begin{center}
\begin{tabular}{|c|c|c|c|c|c|}
\hline
Ordering               & $|\lambda|_{\rm min}$ 
                       & $|\lambda|_{\rm max}$
                       & ${\cal C}$ & $n_{\rm iter}$ \\ 
\hline
\hline
 None                  &    0.02306(77)      &   2.02145(11)
                       &  99.5(3.9) &  501(11) \\
 eo-only               &    0.03511(94)      &   1.56235(76)
                       &  48.1(1.4) &  202(5) \\
\hline
   global              &    0.0838(21)       &   5.906(43)
                       & 76.1(2.5)  &  {\bf 31.2(1)} \\
   local (A)           & 0.0830(23)          &   1.85347(81)
                       & 24.5(1.0)  &  68.0(5) \\
   local (B)           & 0.0900(24)          &   1.5719(10)
                       & {\bf 18.82(55)}  &  76.9(7) \\
\hline
\end{tabular}
\capt{Solver and HMC algorithm performance at $\beta=4.0, 
\kappa=0.26$ on a $32\times 32$ lattice for various eo-ILU schemes.
  \label{tab:eoilu-spectral}}
\end{center}
\end{table}

Table~\ref{tab:eoilu-spectral} presents the eigenvalues of the hermitian 
eo-ILU preconditioned matrix, $\bar{Q}_{\rm ee}$ with the largest and smallest
absolute values, along with the condition number and average number of BiCGStab
solver iterations required for matrix inversion. The same pattern emerges as 
with the one-level ILU scheme comparisons. The best ordering scheme to 
accelerate the BiCGStab algorithm is the global-lexicographic scheme, while 
the scheme with the smallest condition number is one of the local four-flavour 
preconditionings. These optimal schemes are highlighted bold in 
Table~\ref{tab:eoilu-spectral}. The two-level preconditioning scheme reduces 
the condition number of the matrix and the number of iterations required for 
inverting the fermion matrix even further than the single ILU scheme. The 
optimal ordering for matrix inversion out-performs the unpreconditioned matrix 
by a factor of 16, while the optimal scheme for improving the condition number 
reduces this number by a factor of five. 

\subsection{eo-ILU HMC performance}

The eo-ILU preconditioned pseudofermion action was 
tested in a set of HMC algorithm simulations and compared to the
unpreconditioned and even-odd schemes for two fermion masses. 
The studies of ILU HMC in Sec.~\ref{sec:ILU-HMC-perf} found the
benefits of preconditioning were rather weakly dependent on the site ordering, 
but that the schemes leading to the highest HMC acceptance rate were the 
local orderings, such as the familiar even-odd method. 
The lowest condition number was also found to be an indicator of the best 
acceptance rate. Based on this and the results in 
Table~\ref{tab:eoilu-spectral}, local eo-ILU 
ordering B was used. One simulation using the A ordering was also considered, 
and found to give a slightly poorer performance than scheme B. These 
simulations were performed on $64\times 64$ lattices at $\beta=4.0$ and two 
$\kappa$ values, 0.2570 and 0.2605. These parameters corresponded to 
pseudoscalar meson masses of $a m_P = 0.210(3)$ and $a m_P = 0.124(5)$ 
respectively.

\begin{figure}[h!]
  \begin{center}
    \begin{minipage}[h]{12cm} 
      \setlength{\epsfxsize}{10cm}
      \centerline{\epsfbox{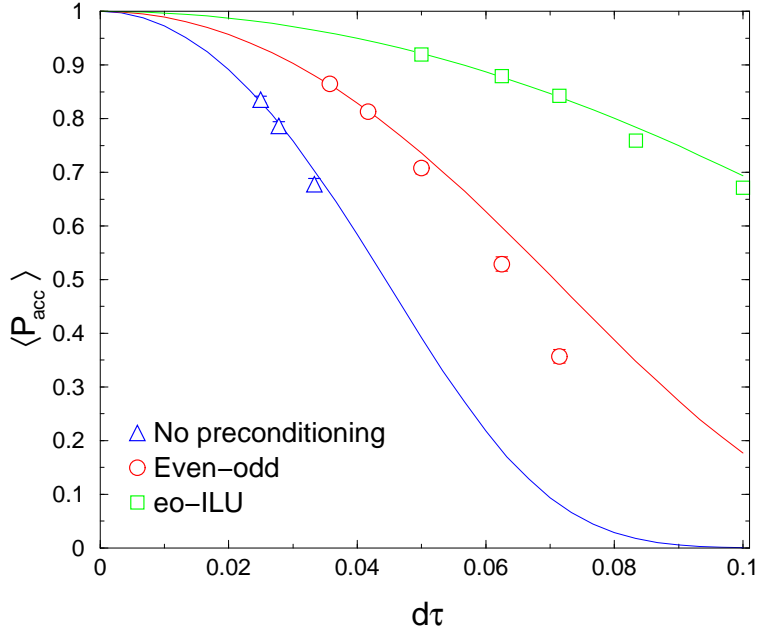}}
      \caption{
        Acceptance rate {\it vs.} molecular dynamics step-size for three
        preconditioning schemes. Dashed lines are fits to 
        Eqn.~(\protect{\ref{eqn:fit}}). All simulations are performed on 
        $64\times 64$ lattices at $\beta=4.0, \kappa=0.2570$ \label{fig:k2570}}
    \end{minipage}
  \end{center}
\end{figure}
\begin{figure}[h!]
  \begin{center}
    \begin{minipage}[h]{12cm} 
      \setlength{\epsfxsize}{10cm}
      \centerline{\epsfbox{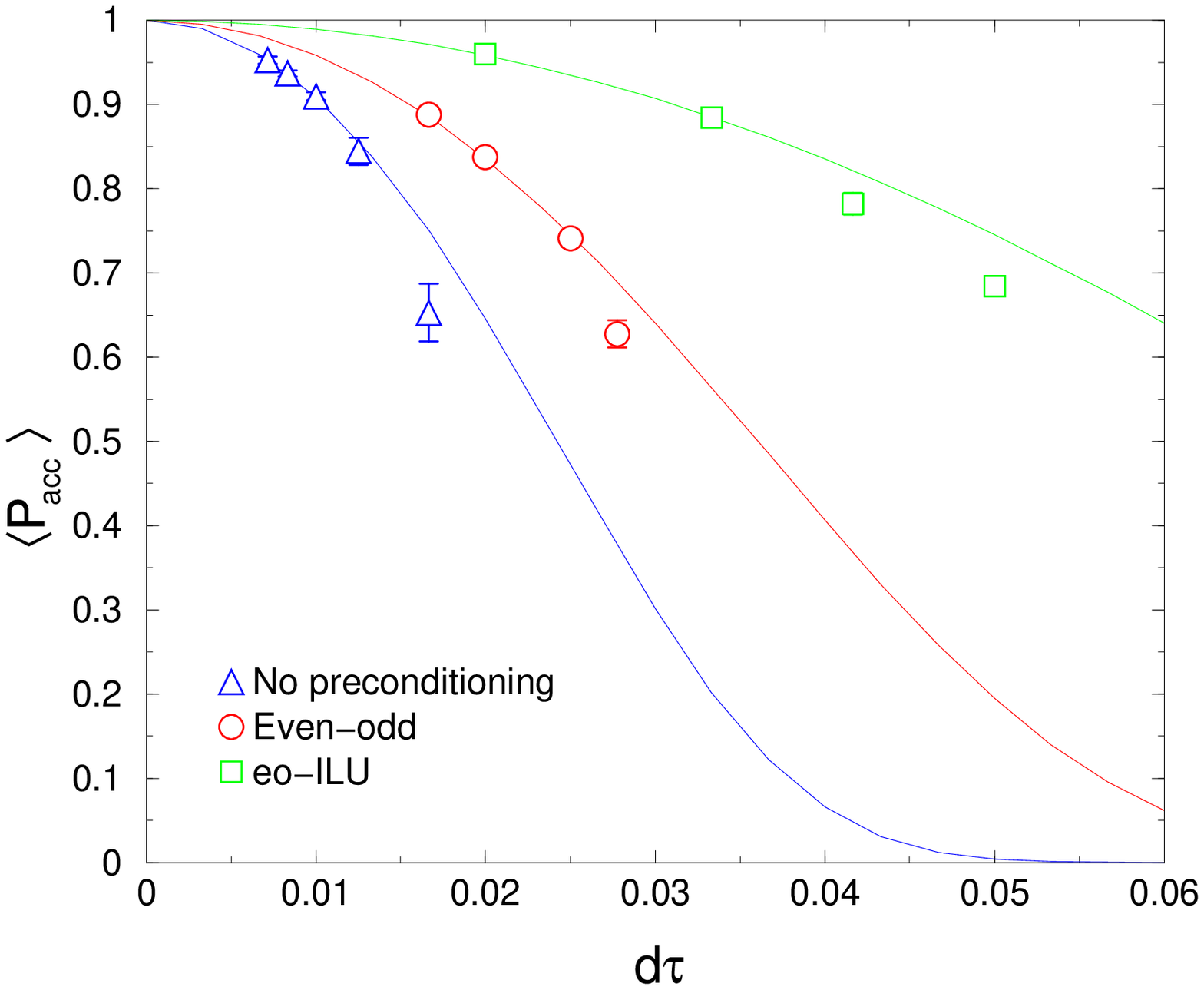}}
      \caption{
        Acceptance rate {\it vs.} molecular dynamics step-size for three
        preconditioning schemes. Dashed lines are fits to 
        Eqn.~(\protect{\ref{eqn:fit}}). All simulations are performed on 
        $64\times 64$ lattices at $\beta=4.0, \kappa=0.2605$ \label{fig:k2605}}
    \end{minipage}
  \end{center}
\end{figure}
The dependence
of the acceptance probability on step-size for the two fermion masses are
presented in Figs.~\ref{fig:k2570} and \ref{fig:k2605}. In these figures, fits
to the expected small-step-size behaviour of the acceptance probability
\cite{Gupta:1990ka} are included,
\begin{equation}
 \langle\pacc\rangle = \mbox{erfc }\left(\frac{d\tau}{\tau_0}\right)^2,
  \label{eqn:fit}
\end{equation}
where $\tau_0$ is a fit parameter and determines the characteristic scale of the
equations of motion. 
$\tau_0$ presents a reliable estimate of the algorithm efficiencies, 
assuming the autocorrelations along the Markov chains for the different
preconditionings are the same at a fixed Metropolis acceptance. While these 
autocorrelations have not been studied in detail, this criterion does appear to
hold approximately. 
Eqn.~(\ref{eqn:fit}) models the expected acceptance rate only in
the limit $d\tau \rightarrow 0$. The fit gives an unacceptable $\chi^2$ once 
the acceptance rate falls below approximately 80\%.
At the lighter fermion mass $(\kappa=0.2605)$ the acceptance rate tended to 
break down rather suddenly as $d\tau$ was increased. The UKQCD collaboration 
\cite{Joo:2000dh} studied this phenomena in detail, and concludes that the 
leap-frog integrator becomes unstable at a critical value of $d\tau$, which
decreases as the light quark mass is reduced. 
To determine $\tau_0$ from a fit, the number of data points 
included was increased (starting with the smallest $d\tau$) until 
$\chi^2/N_{\rm df} > 1.5$.
These fits are summarised in Table~\ref{tab:tau-0}, along with 
the masses of the lightest mesons. Table~\ref{tab:tau-0} also includes the
ratio of the characteristic molecular-dynamics time-scale of unpreconditioned
HMC and the even-odd and eo-ILU schemes. 
For both values of the hopping parameter, the eo-ILU
algorithm characteristic time is larger than standard HMC by a factor of about 
three and the even-odd method by about two. 
No significant mass dependence on this improvement is seen, although only two
fermion masses were simulated.

\begin{table}[!ht]
\begin{center}
\begin{tabular}{|c|c|c|c|c|}
\hline
 \multicolumn{5}{|c|}{$\beta = 4.0,\;\kappa = 0.2570$
                      \hspace{3ex}$(a m_P = 0.210 \pm 0.003)$ } \\
\hline
 Preconditioning & $n_{\rm fit}$ & $\chi^2/N_{\rm df}$ & $\tau_0$  &
                                                            Improvement \\
\hline
    none         &   2           & 1.00                & 0.0643(9) & --- \\
    eo           &   2           & 0.16                & 0.102(1)  & 1.59(3) \\
    eo-ILU (B)   &   3           & 0.26                & 0.189(2)  & 2.94(5) \\
\hline
\hline
 \multicolumn{5}{|c|}{$\beta = 4.0,\;\kappa = 0.2605$
                      \hspace{3ex}$(a m_P = 0.124\pm 0.005)$ } \\
\hline
 Preconditioning & $n_{\rm fit}$ & $\chi^2/N_{\rm df}$ & $\tau_0$  &
                                                            Improvement \\
\hline
    none         &   4           &     0.26            & 0.0351(6) & --- \\ 
    eo           &   3           &     0.23            & 0.0522(6) & 1.48(3) \\
    eo-ILU (B)   &   2           &     0.11            & 0.104(2)  & 2.96(8) \\
\hline
\end{tabular}
\capt{Fits to Eqn.~(\protect{\ref{eqn:fit}}) at two $\kappa$ values for three 
HMC algorithms with no preconditioning, one step (even-odd) and two step 
(eo-ILU) preconditioning.  $n_{\rm fit}$ indicates the number of points 
(starting from the lowest $d\tau$ data) that could be included in the fit
before an unacceptably high $\chi^2/N_{\rm df}$ was found. 
  \label{tab:tau-0}}
\end{center}
\end{table}

\section{Discussion: further implementations \label{sec:implementations}}

The preconditioned pseudofermion method, tested in the Schwinger model, can be
applied directly in simulations of 4d gauge theories, such as QCD. As present,
these computationally demanding calculations are being performed on large
parallel computers. 
In Sec.~\ref{sec:simulations}, some correlation was seen between the
performance of the preconditioned HMC algorithm and the condition number of the
hermitian fermion matrix, $Q$. Finding the optimal site ordering is then
reduced to minimising this condition number and a
direct comparison of HMC performance from costly simulations can be avoided.
Only a small set of orderings was tested in the two-dimensional study; the
benefits of a given ordering will be dependent on the precise properties of the 
matrix in question and will differ in two and four dimensions.

These simulations also demonstrated that a significant 
enhancement in the performance of the HMC algorithm was found by using a 
two-step (eo-ILU) preconditioning of the pseudofermions. A highly efficient
ordering for increasing the acceptance of the Metropolis test is a local one; 
{\it ie.} matrix operations still involve only a small neighbourhood
(out to two hops) around each site and do not require global lattice sweeps for
their operation. This means these preconditionings can be
applied to simulations on parallel computers, as many processors can be
working on independent portions of the local eo-ILU matrix-vector operation. 
The four-dimensional lattice would required sixteen indices to cover the
orderings of the even-odd matrix and it is unclear whether this would lead to 
prohibitively expensive or complex communications on parallel computers. A 
direct test seems the only way to assess this. Sec.~\ref{sec:simulations} 
determined that the acceptance rate is not critically dependent on the ordering 
and a better scheme for efficient parallelism could be found. 

Ref.~\cite{Jansen:1997yt} demonstrated that the Sheikholeslami-Wohlert (SW) 
action can be simulated using even-odd preconditioned HMC. 
As with the Wilson matrix, the even-odd SW matrix can be ILU preconditioned 
again, leading to an eo-ILU scheme with a possible faster inverter 
performance and a larger useful molecular-dynamics step-size. Many large-scale 
dynamical quark simulations ({\it eg.} CP-PACS, UKQCD; see 
Ref.~\cite{Mawhinney:2000fw} for a review) use the SW fermion formulation. 
Lexicographic preconditioning has also been applied to more complex fermion 
matrices \cite{Bietenholz:1999wx}, and can be extended to highly improved 
actions with interactions beyond nearest neighbours, such as the 
Symanzik-improved $D234$ action \cite{Alford:1998yy} and fixed point actions 
\cite{Bietenholz:1996cy,DeGrand:1998pr}. This work suggests HMC simulations of
these fermion actions can be accelerated, even though the standard even-odd
decomposition does not decouple even and odd sub-lattices. 
Note also that other dynamical fermion methods, such as the 
``Kentucky algorithm'' \cite{Lin:2000qu} may also be enhanced by similar ILU or
eo-ILU preconditioning.

As a final remark here, it is worth examining the hopping parameter 
expansion of the inverse preconditioned matrices. The original Wilson matrix 
inverse has a term at ${\cal O}(\kappa)$ while after any ILU preconditioning, 
the leading term appears at ${\cal O}(\kappa^2)$. With the two-step eo-ILU 
preconditioning, the lowest order term is now ${\cal O}(\kappa^4)$.

\section{Conclusions}

In this paper, ILU matrix preconditioning has been applied directly to the
pseudofermion action of the hybrid Monte Carlo algorithm. 
The optimal ordering scheme for a single level of preconditioning 
in the pseudofermion action are local, like the simple even-odd 
checkerboard, while for inverting 
the fermion matrix a global-lexicographic ordering is best. The performance of
the HMC algorithm was seen to depend weakly on the ordering, but to correlate 
with the condition number of the fermion matrix, while inversion performance 
has a more complex behaviour and is influenced strongly by the site indexing. 

A two-level
scheme was then introduced, in which the even-odd fermion matrix on a
single sub-lattice was subsequently ILU preconditioned (eo-ILU). 
The global lexicographic ordering for this scheme was proposed in Section 4 of
Ref.~\cite{deForcrand:1997ck}.  In direct analogy, the
optimal ordering on the sub-lattice for inversion was found to be a global
lexicographic scheme, while the HMC algorithm performed best with a more local,
``four-flavour'' decomposition. This optimal two-level scheme was found to 
improve the performance of the HMC algorithm by a factor of two over
even-odd pseudofermions and a factor of three relative to 
the unmodified HMC method.


\section*{Acknowledgements}

This work was funded by the US DOE under grant No. DE-FG03-97ER40546.
I am grateful to Julius Kuti for a critical reading of this manuscript. 
I would like to thank Philippe de Forcrand for bringing the work of 
Ref.~\cite{deForcrand:1997ck} to my attention. 



\begin{thebibliography}{99}

\bibitem{Duane:1987de}
S.~Duane, A.~D.~Kennedy, B.~J.~Pendleton and D.~Roweth,
Phys.\ Lett.\  {\bf B195}, 216 (1987).

\bibitem{Fischer:1996th}
S.~Fischer, A.~Frommer, U.~Glassner, T.~Lippert, G.~Ritzenhofer and
K.~Schilling,
Comput.\ Phys.\ Commun.\  {\bf 98}, 20 (1996) [hep-lat/9602019].

\bibitem{Gupta:1989kx}
R.~Gupta, A.~Patel, C.~F.~Baillie, G.~Guralnik, G.~W.~Kilcup and S.~R.~Sharpe,
Phys.\ Rev.\  {\bf D40} (1989) 2072.

\bibitem{deForcrand:1997ck}
P.~de Forcrand and T.~Takaishi,
Nucl.\ Phys.\ Proc.\ Suppl.\  {\bf 53} (1997) 968 [hep-lat/9608093].

\bibitem{Oyanagi}
Y.~Oyanagi,
Comput.\ Phys.\ Commun.\  {\bf 42}, 333 (1986).

\bibitem{Lippert:1998qy}
T.~Lippert {\it et al.}  [SESAM Collaboration],
Nucl.\ Phys.\ Proc.\ Suppl.\  {\bf 63}, 946 (1998) [hep-lat/9712020].

\bibitem{Eisenstat}
S.~Eisenstat.
J.\ Sci.\ Stat.\ Comput.\ 2 1 (1981). 

\bibitem{Sexton:1992nu}
J.~C.~Sexton and D.~H.~Weingarten,
Nucl.\ Phys.\  {\bf B380}, 665 (1992).

\bibitem{Gupta:1990ka}
S.~Gupta, A.~Irback, F.~Karsch and B.~Petersson,
Phys.\ Lett.\  {\bf B242} (1990) 437.

\bibitem{Joo:2000dh}
B.~Joo, B.~Pendleton, A.~D.~Kennedy, A.~C.~Irving, J.~C.~Sexton, S.~M.~Pickles
and S.~P.~Booth [UKQCD Collaboration],
hep-lat/0005023.

\bibitem{Jansen:1997yt}
K.~Jansen and C.~Liu,
Comput.\ Phys.\ Commun.\  {\bf 99} (1997) 221
[hep-lat/9603008].

\bibitem{Mawhinney:2000fw}
R.~D.~Mawhinney,
Nucl.\ Phys.\ Proc.\ Suppl.\  {\bf 83-84} (2000) 57
[hep-lat/0001032].

\bibitem{Bietenholz:1999wx}
W.~Bietenholz, N.~Eicker, A.~Frommer, T.~Lippert, B.~Medeke, K.~Schilling and
G.~Weuffen,
Comput.\ Phys.\ Commun.\  {\bf 119}, 1 (1999) [hep-lat/9807013].

\bibitem{Alford:1998yy}
M.~Alford, T.~R.~Klassen and G.~P.~Lepage,
Phys.\ Rev.\  {\bf D58} (1998) 034503
[hep-lat/9712005].

\bibitem{Bietenholz:1996cy}
W.~Bietenholz and U.~J.~Wiese,
Nucl.\ Phys.\  {\bf B464} (1996) 319
[hep-lat/9510026].

\bibitem{DeGrand:1998pr}
T.~DeGrand  [MILC Collaboration],
Phys.\ Rev.\  {\bf D58} (1998) 094503
[hep-lat/9802012].

\bibitem{Lin:2000qu}
L.~Lin, K.~F.~Liu and J.~Sloan,
Phys.\ Rev.\  {\bf D61}, 074505 (2000) [hep-lat/9905033].


\end{thebibliography}
\end{document}